\documentstyle[aps,epsf,preprint]{revtex}
\begin{document}
\draft
\title{Search for $a_0(980)$ and $f_0(980)$ mesons in  $\phi\to\gamma K\bar K$.}
\author
{N.N. Achasov
\thanks{achasov@math.nsc.ru}
 and V.V. Gubin
\thanks{gubin@math.nsc.ru}
  \\ Laboratory of Theoretical Physics,
\\ Sobolev Institute for Mathematics \\ 630090, Novosibirsk-90,
Russia}
\date{\today}
\maketitle
\begin{abstract}
 We study the decay $\phi\to\gamma K^+K^-$
 taking into account the scalar mesons production
 $\phi\to\gamma(f_0+a_0)\to\gamma K^+K^-$ and the
 final state radiation. We note that
 the relative sign between the final state radiation amplitude
 and the scalar meson production amplitude is fixed in the
$K^+K^-$ loop model which describes data on the $\phi\to \gamma
f_0\to\gamma\pi\pi$ and $\phi\to\gamma a_0\to\gamma\pi\eta$
decays. As consequence this model predicts the definite
interference between the final state radiation and the scalar
resonance production amplitudes.  We calculate the mass spectra of
the $\phi\to\gamma K^+K^-$ decay and the
 differential cross-sections for $e^+e^-\to\phi\to\gamma
K^+K^-$ and for $e^+e^-\to\phi\to\gamma K^0\bar K^0$ reactions.
\end{abstract}

\section{Introduction}

As  was shown in a number of papers, see Refs.
\cite{achasov-89,close-93,nutral,shevchenko,lucio,phase} and
references therein, the study of the radiative decays
$\phi\to\gamma a_0\to\gamma\pi\eta$ and $\phi\to\gamma f_0\to
\gamma\pi\pi$ can shed light on the problem of the scalar
$a_0(980)$ and $f_0(980)$ mesons. These decays have been studied
not only theoretically but also experimentally. Present time data
 have already been  obtained from Novosibirsk with the detectors SND
 \cite{snd-1,snd-2,snd-fit,snd-ivan} and CMD-2 \cite{cmd},
 which give the following branching ratios :
$BR(\phi\to\gamma\pi\eta)=(0.88\pm0.14\pm0.09)\cdot10^{-4}$
\cite{snd-fit}, $BR(\phi\to\gamma\pi^0\pi^0)=
(1.221\pm0.098\pm0.061)\cdot10^{-4}$ \cite{snd-ivan} and
$BR(\phi\to\gamma\pi\eta)=(0.9\pm0.24\pm0.1)\cdot10^{-4}$,
$BR(\phi\to\gamma\pi^0\pi^0)=(0.92\pm0.08\pm0.06)\cdot10^{-4}$
\cite{cmd}.

\begin{figure}
 \centerline{ \epsfxsize=12cm \epsfysize=4cm
\epsfbox{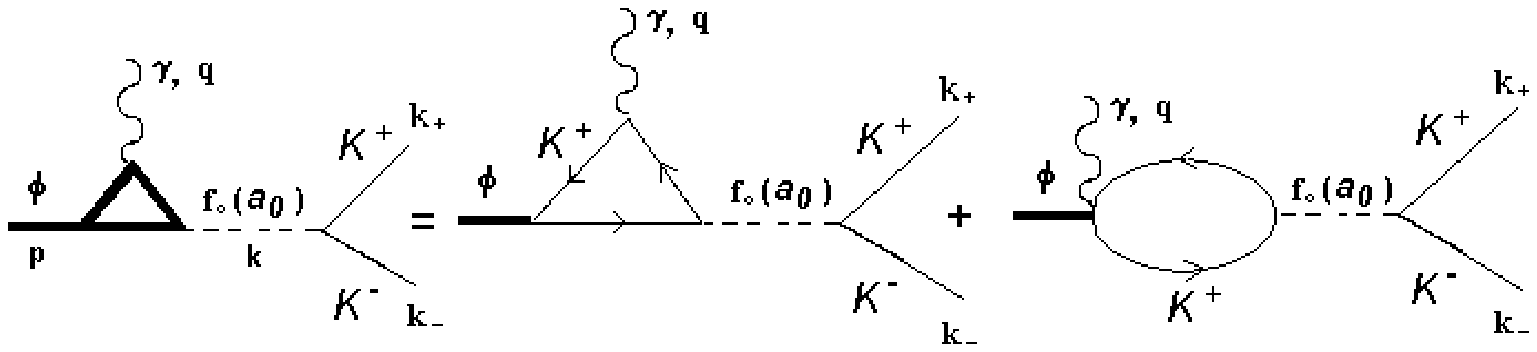}}
 \caption{Diagrams of the $K^+K^-$ loop model.  }
 \label{loop}
\end{figure}

 These data give evidence   in favor of the four-quark $(q^2\bar
q^2)$
\cite{achasov-89,nutral,jaffe,ach-84,ach-91,ach-98,black,achasov-01}
nature of the scalar $a_0(980)$ and $f_0(980)$ mesons, and in
favor of  the one-loop mechanism $\phi\to K^+K^-\to\gamma a_0$ and
$\phi\to K^+K^-\to\gamma f_0$, suggested in Ref.\cite{achasov-89},
see Fig.\ref{loop}.

\begin{figure}
 \centerline{\epsfxsize=12cm \epsfysize=3cm
\epsfbox{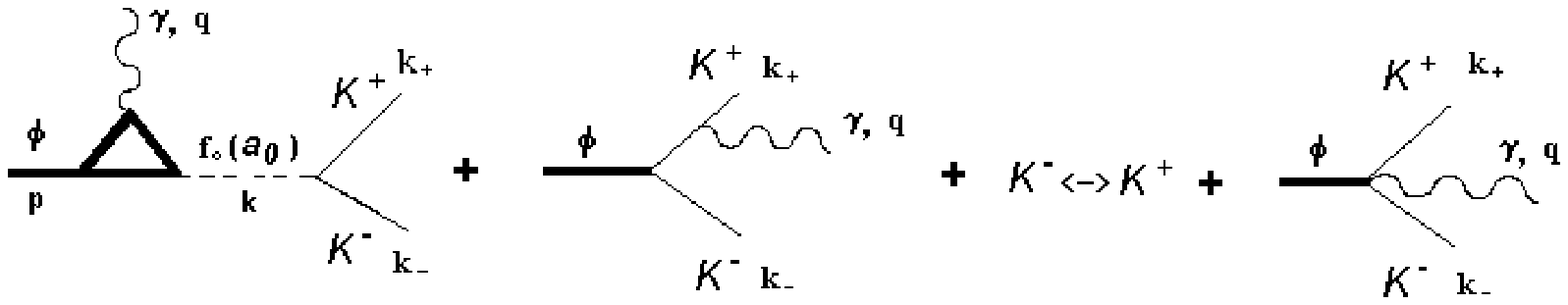}}
 \caption{Diagrams of the signal $\phi\to\gamma K^+K^-$ and background processes.}
 \label{diagrams-phi}
\end{figure}

It is clear that the  relative sign between the final state
radiation amplitude,  and the $\phi\to\gamma(a_0+f_0)\to\gamma
K^+K^-$ amplitude in the $K^+K^-$ loop model is fixed , see Figs.
\ref{loop} and \ref{diagrams-phi}. As a consequence this model
predicts the definite interference between the final state
radiation amplitude and the scalar resonance production.
 That is why the study of the $\phi\to\gamma K^+K^-$ decay is very important.
 Such an investigation requires high statistics. It can be carried
 out in the $\gamma N\to\phi N(\Delta)\to \gamma K^+K^-N(\Delta)$
 reactions at Jefferson Laboratory and in the
 $e^+e^-\to\phi\to\gamma K^+K^-$ reaction at DA$\Phi$NE.
 Note that if the $f_0$ and $\sigma$ mesons mixing is taken into
 account in the $\phi\to\gamma(a_0+f_0(\sigma))\to\gamma
K^+K^-$ reaction then the interference sign is not fixed. But the
influence of the $\sigma$ meson on the signal is small and it is
considered as a correction to the interference term.

 We calculate the $\phi\to\gamma K^+K^-$ spectra and the
differential cross-sections for $e^+e^-\to\phi\to\gamma K^+K^-$,
using parameters of $a_0(980)$ and $f_0(980)$ mesons from
Ref.\cite{achasov-01,snd-fit,snd-ivan}
 which describe the $\phi\to\gamma
a_0\to\gamma\pi\eta$ and $\phi\to\gamma f_0\to\gamma\pi\pi$
decays, and show that the contribution of the interference between
the scalar resonance production and the final state radiation is
order of magnitude more than the contribution of the scalar
resonance production. That is the reason that today's facilities
give the good capabilities to investigate our issue.

We also calculate the differential cross section for
$e^+e^-\to\phi\to\gamma K^0\bar K^0$ which is very important for
the determination of the background affecting the precision
measurements of CP violation.

The paper is organized as follows.

All needed formulae are considered in Sec. II. In Sec III the
calculations are carried out and the obtained results are
discussed. A brief summary is given in Conclusion. Notice that
Ref. \cite{lucio-2} was dedicated to the $e^+e^-\to\phi\to\gamma
K\bar K$ reactions at the $\phi$ meson region. Unfortunately, the
authors of Ref. \cite{lucio-2} considered mistakenly  that the
relative sign between the final state radiation amplitude and the
$e^+e^-\to\gamma(a_0+f_0)\to\gamma K^+K^-$ amplitude is not fixed
in the $K^+K^-$ loop model. In addition, they took parameters of
$a_0(980)$ and $f_0(980)$ mesons a priori to a large measure.

\section{Model}

Let us consider the production of the scalar $R=a_0,f_0$ meson
through the loop of the charged $K$ mesons, $\phi\to
K^+K^-\to\gamma R$, see \cite{achasov-89,achasov-95}. The diagrams
are presented in Fig.\ref{loop}. The production amplitude
 $\phi\to\gamma R$  in the rest frame of the $\phi$ meson is:
\begin{equation}
M=g_{RK^+K^-}g(m^2)\vec e(\phi)\vec e(\gamma)
\end{equation}
where $t=m^2=k^2$, $\vec e(\phi)$ and $\vec e(\gamma)$ are the
polarization vectors of the $\phi$ meson and the photon
respectively. The expressions for $g(m^2)$ were obtained in the
point-like particle model, which is adequate to the compact
hadrons: the four-quark $(q^2\bar q^2)$ or $q\bar q$ states
\cite{achasov-89} and in the extended  scalar $K\bar K$ molecule
model \cite{shevchenko}. The amplitude of the decay
$\phi\to(\gamma f_0+\gamma a_0)\to\gamma K^+K^-$, i.e., the signal
amplitude,  is
\begin{equation}
M_s=M_s^{\mu}e^{\mu}(\phi)=\left (
\frac{g_{f_0 K^+K^-}^2}{D_{f_0}(m)}+\frac{g_{a_0
K^+K^-}^2}{D_{a_0}(m)}\right )
g(m^2)(q^{\mu}\frac{e(\gamma)p}{pq}-e^{\mu}(\gamma))e^{\mu}(\phi)
\label{amplituda-0}
\end{equation}
where $s=p^2$.
The mass spectrum:
\begin{equation}
\frac{d\Gamma_s}{dm}= \frac{m}{(2\pi)^3 36s\sqrt{s}}\left
|g(m^2)\left(\frac{g_{f_0 K^+K^-}^2}{D_{f_0}(t)}+\frac{g_{a_0
K^+K^-}^2}{D_{a_0}(t)}\right)\right
|^2(s-t)b\sqrt{1-\frac{4m_{K^+}^2}{m^2}} \label{spector-0}
\end{equation}
We introduce symmetrical angle cut $-b\leq\cos\theta_{K\gamma}\leq
b$, where $\theta_{K\gamma}$ is the angle between the photon and
the $K^+$ meson momenta in the dikaon rest frame. We consider also
the  $f_0$ and $\sigma$ mesons mixing. In this case the amplitude
is
\begin{equation}
M_s=\left (
g_{RK+K^-}\sum_{RR'}G_{RR^\prime}^{-1}(m)g_{R'K^+K^-}+\frac{g_{a_0
K^+K^-}^2}{D_{a_0}(m)}\right )
g(m^2)(q^{\mu}\frac{e(\gamma)p}{pq}-e(\gamma)^{\mu})e^{\mu}(\phi)
\label{amplituda}
\end{equation}
where we take into account the mixing of
the $f_0$ and $\sigma$ mesons, $R,R'=f_0,\sigma$ by means of the
 matrix of the inverse propagators $G(m)$.
The all necessary expressions could be found in \cite{nutral}.
The mass spectrum in this case is
\begin{equation}
\frac{d\Gamma_s}{dm}= \frac{m}{(2\pi)^3 36s\sqrt{s}}\left
|g(m^2)\left( g_{R
K^+K^-}\sum_{RR'}G_{RR^\prime}^{-1}(m)g_{R'K^+K^-}+\frac{g_{a_0
K^+K^-}^2}{D_{a_0}(m)}\right)\right
|^2(s-t)b\sqrt{1-\frac{4m_{K^+}^2}{m^2}} \label{spector}
\end{equation}


Let us consider the background related to the final state
radiation, see Fig.\ref{diagrams-phi}.
 The amplitude of the process
is
\begin{eqnarray}
\label{fon-0} &&M_{fin}=2eg_{\phi K^+K^-}T^{\mu}e^{\mu}(\phi), \\ \nonumber
&&T^{\mu}=\frac{e(\gamma)k_-}{qk_-}(k_+-\frac{p}{2})^{\mu}+
\frac{e(\gamma)k_+}{qk_+}(k_--\frac{p}{2})^{\mu}+e(\gamma)^{\mu}
\end{eqnarray}
The mass spectrum is
\begin{equation}
\frac{d\Gamma_{fin}}{dm}=\frac{\alpha mg_{\phi
K^+K^-}^2}{12\pi^2\sqrt{s}}\left(\frac{v(x)}{x}\Biggl(x^2-(1-\xi)(1-x)\Biggr)+
 (1-\xi)(1-x-\frac{\xi}{2})\frac{1}{x}\ln\frac{1+v(x)}{1-v(x)}\right)
\end{equation}
where $v(x)=b\sqrt{1-\frac{\xi}{1-x}}$. We identify
$\xi=\frac{4m_{K^+}^2}{s}$ and $x=\frac{2\omega}{\sqrt{s}}$,
$t=s(1-x)=m^2$ for convenience.

The interference between the amplitudes (\ref{amplituda-0}) and
(\ref{fon-0}) gives the expression
\begin{eqnarray}
&&\frac{d\Gamma_{int}}{dm}=\frac{\alpha mg_{\phi
K^+K^-}}{\sqrt{4\pi\alpha s}}Re\Biggl[g(m^2)\left(\frac{g_{f_0
K^+K^-}^2}{D_{f_0}(m)} + \frac{g_{a_0
K^+K^-}^2}{D_{a_0}(m)}\right)\Biggr]\times
\\ \nonumber
&&\times\Biggl\{v(x)+\frac{\xi}{2}\ln\frac{1-v(x)}{1+v(x)}\Biggr\}
\end{eqnarray}

For the decay $\phi\to\gamma(f_0+a_0)\to\gamma K^0\bar K^0$, due
to the destructive interference between the $a_0$ and $f_0$
mesons, we have the following expressions
\begin{equation}
\frac{d\Gamma_s}{dm}= \frac{m}{(2\pi)^3 36s\sqrt{s}}\left
|g(m^2)\left(\frac{g_{f_0 K^0\bar K^0}^2}{D_{f_0}(t)}-\frac{g_{a_0
K^0\bar K^0}^2}{D_{a_0}(t)}\right)\right
|^2(s-t)b\sqrt{1-\frac{4m_{K^0}^2}{m^2}} \label{spector-K0}
\end{equation}

\begin{equation}
\frac{d\Gamma_s}{dm}= \frac{m}{(2\pi)^3 36s\sqrt{s}}\left
|g(m^2)\left( g_{R K^0\bar
K^0}\sum_{RR'}G_{RR^\prime}^{-1}(m)g_{R'K^0\bar K^0}-\frac{g_{a_0
K^0\bar K^0}^2}{D_{a_0}(m)}\right)\right
|^2(s-t)b\sqrt{1-\frac{4m_{K^0}^2}{m^2}}
\end{equation}

\begin{figure}
 \centerline{\epsfxsize=12cm \epsfysize=8cm
\epsfbox{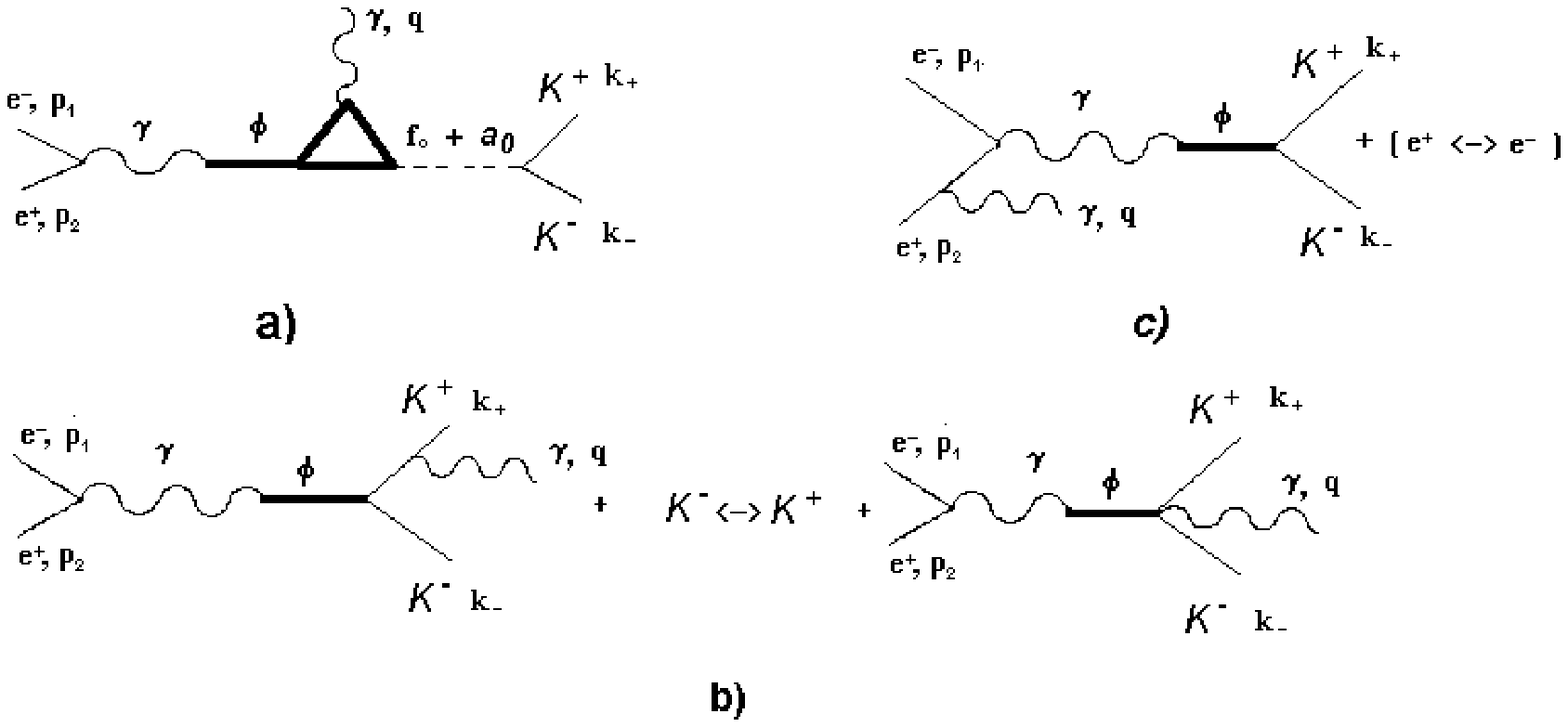}}
 \caption{Diagrams of the signal $e^+e^-\to\phi\to\gamma K^+K^-$
 and background processes.}
 \label{diagrams-ee}
\end{figure}

Let us consider the reaction $e^+e^-\to\phi\to(\gamma f_0+\gamma
a_0)\to\gamma K^+K^-$, see Fig.\ref{diagrams-ee}. The amplitude of
the reaction is
\begin{equation}
M_{ee}= e\bar
u\gamma^{\mu}u\frac{em_{\phi}^2}{f_{\phi}sD_{\phi}(s)}M_s^{\mu}
\label{amplituda-ee}
\end{equation}
The differential cross-section is
\begin{equation}
\label{ee-signal}
 \frac{d\sigma_{\phi}}{dm}=\frac{\alpha^2}{24\pi s^3}
\left(\frac{m_{\phi}^2}{f_{\phi}}\right)^2\frac{m}{|D_{\phi}(s)|^2}
\left |g(m^2)\left( \frac{g_{f_0 K^+K^-}^2}{D_{f_0}(m)}+
\frac{g_{a_0 K^+K^-}^2}{D_{a_0}(m)}\right)\right |^2
(s-t)\sqrt{1-\frac{4m_{K^+}^2}{m^2}}\quad (a+\frac{a^3}{3})b
\end{equation}
The coupling
constants  $g_{f_0K^+K^-}$ and $f_{\phi}$ are related to the
widths in the following way:

\begin{equation}
\Gamma(f_0\to K^+K^-,m)=
\frac{g_{f_0K^+K^-}^2\sqrt{m^2-4m_{K^+}^2}}{24\pi m^2},
\qquad\Gamma(V\to
e^+e^-,s)=\frac{4\pi\alpha^2}{3}(\frac{m_V^2}{f_V})^2
\frac{1}{s\sqrt{s}}. \label{pipi}
\end{equation}
We introduce two symmetrical angle cuts:
$-a\leq\cos\theta_{\gamma}\leq a$, where $\theta_{\gamma}$ is the
angle between the photon momentum and the electron beam in the
center of mass frame of the reaction under consideration  and
$-b\leq\cos\theta_{K\gamma}\leq b$, where $\theta_{K\gamma}$ is
the angle between the photon and the $K^+$ meson momenta in the
dikaon rest frame. Notice that the kaons are slow in our case,
$\sqrt{1-4m_K^2/m^2}\leq0.25$, which is why $\theta_{K\gamma}$
angle cut practically does not change the results (at
$150^{\circ}>\theta_{K\gamma}>30^{\circ}$ changes are within 13\%)
in contrast to the pion production \cite{phase,charge}.

The basic background to the process under study has came from the
initial electron radiation  ( see Fig.\ref{diagrams-ee}c ) and the
radiation from the final kaons ( Fig.\ref{diagrams-ee}b ). The
initial state radiation does not interfere with the final state
radiation and with the signal in the differential cross section
integrated over all angles since the charged kaons are in the C=-1
state. This is true also when the angle cuts are symmetrical.

Introducing the symmetrical angle cuts  considerably decreases the
background from the initial state radiation because of the photons
in this case are emitted along the beams mainly.

Let us consider the background related to the final state
radiation. The amplitude of the process is
\begin{eqnarray}
\label{fon} &&M_{fin}=e^2\bar
u\gamma^{\mu}u\frac{em_{\phi}^2}{f_{\phi}}\frac{1}
{sD_{\phi}(s)}2g_{\phi K^+K^-}T^{\mu}.
\end{eqnarray}

It is convenient to give the differential cross section in the
form \cite{phase,charge}:
\begin{eqnarray}
\label{fonrho}
&&\frac{d\sigma_f}{dm}=2\sigma_0(s)\frac{m}{s}F(x,a,b)
\\ \nonumber &&
F(x,a,b)=\frac{2\alpha}{\pi(1-\xi)^{3/2}}\Biggl\{\frac{3}{2}(a-
 \frac{a^3}{3})F_1+\frac{3}{4}a(1-a^2)F_2\Biggr\} \\ \nonumber
&&
F_1=\frac{1}{x}\Biggl(x^2-\frac{\xi(1-\xi)(1-x)}{(1-b^2)(1-x)+b^2\xi}
\Biggr)f(x)+
 (1-\xi)(1-x-\frac{\xi}{2})\frac{1}{x}\ln\frac{1+f(x)}{1-f(x)} \\ \nonumber
&&
F_2=\frac{1}{x}\Biggl(\frac{\xi^2(x-1)}{(1-b^2)(1-x)+b^2\xi}+2x-2-x^2
\Biggr)f(x)+
 \xi(2-x-\frac{\xi}{2})\frac{1}{x}\ln\frac{1+f(x)}{1-f(x)} \nonumber
\end{eqnarray}
where $f(x)=b\sqrt{1-\frac{\xi}{1-x}}$. The nonradiative cross
section $e^+e^-\to K^+K^-$ is:
\begin{equation}
\sigma_0(s)=\frac{\pi\alpha^2}{3s}(1-\xi)^{3/2}|F_{\phi}(s)|^2.
\end{equation}
where
 $|F_{\phi}(s)|^2=(\frac{g_{\phi K^+K^-}}{f_{\phi}})^2\frac{m_{\phi}^4}
{|D_{\phi}(s)|^2}$.

The interference between the amplitudes from Eqs.
(\ref{amplituda-ee}) and (\ref{fon}) is equal \cite{phase,charge}
\begin{eqnarray}
&&\frac{d\sigma_{int}}{dm}=\frac{\alpha^3}{s^2}\left(\frac{
g_{\phi K^+K^-}}{f_{\phi}}\right)
\left(\frac{mm_{\phi}^4}{f_{\phi}\sqrt{4\pi\alpha}
|D_{\phi}|^2}\right) Re\Biggl[\frac{g_{f_0
K^+K^-}^2g(m^2)}{D_{f_0}(m)}+ \frac{g_{a_0
K^+K^-}^2g(m^2)}{D_{a_0}(m)}\Biggr]\times
\\ \nonumber
&&\times\Biggl\{f(x)+\frac{\xi}{2}\ln\frac{1-f(x)}{1+f(x)}\Biggr\}
(a+\frac{a^3}{3})
\end{eqnarray}

In the similar way let us give the expression  for the
differential cross section of the initial state radiation
\cite{phase,charge}.

\begin{eqnarray}
\label{initial} &&\frac{d\sigma_i}{d
m}=2\sigma_0(t)\frac{m}{s}H(x,a,b)
\\
&&H(x,a,b)=\frac{\alpha}{\pi}\Biggl[\Biggl(\frac{2(1-x)+x^2}{x}\ln
\frac{1+a}{1-a}-ax\Biggr)(\frac{3b}{2}-\frac{b^3}{2})+
\frac{3a(1-x)(b^3-b)}{x}\Biggr] \nonumber
\end{eqnarray}
Evaluating $H(x,a,b)$ we ignored the electron mass.

The total cross section of the one photon annihilation with the
soft photon radiation and with the virtual corrections of order
$\alpha$ is given by
\begin{eqnarray}
\label{rad}
&&\sigma(s)=\tilde\sigma(s)\{1+\frac{2\alpha}{\pi}[(L-1)\ln\frac{2
\omega_{min}}{\sqrt{s}}+\frac{3}{4}L+\frac{\pi^2}{6}-1]\}  \\
&&\tilde\sigma(s)=(\sigma_{\phi}(s)+\sigma_{int}(s)+\sigma_{i}+\sigma_{f})
\frac{1}{|1-\Pi(s)|^2}  \nonumber
\end{eqnarray}
where $\omega_{min}$ is the minimal photon energy registered,
$L=\ln\frac{s}{m_e^2}$ is the "main" logarithm.  The electron
vacuum polarization of order $\alpha$ is
\begin{equation}
\Pi(s)=\frac{\alpha}{3\pi}(L-\frac{5}{3})
\end{equation}
where the contribution of muons and light hadrons is ignored.

For the propagator of the $\phi$ meson we use the expression:
\begin{equation}
D_{\phi}(s)=m_{\phi}^2-s-is\frac{g^2_{\phi
K^+K^-}}{48\pi}\Biggl[(1-
\frac{4m_{K^+}^2}{s})^{3/2}+\frac{1}{Z(s)}(1-\frac{4m_{K^0}^2}{s})^{3/2}+0.1p_{\pi\rho}^3\Biggr]
\end{equation}
where $g_{\phi K^+K^-}=4.68$,
$p_{\pi\rho}=\sqrt{(s-(m_{\pi}-m_{\rho})^2)(s-(m_{\pi}+m_{\rho})^2)}/(2\sqrt{s})$,
 and the factor $Z(s)=1+\alpha\pi\frac{1+v^2}{2v}$,
$v=(1-\frac{4m_{K^+}^2}{s})^2$ takes into account the radiative
correction $g_{\phi K^+K^-}/Z(s)=g_{\phi K^0\bar K^0}$, see
details in \cite{kulon} .

For the scalar mesons propogators we use the expressions:
\begin{equation}
\label{propagator} D_R(m^2)=m_R^2-m^2+\sum_{ab}g_{Rab}[Re
P_R^{ab}(m_R^2)-P_R^{ab}(m^2)],
\end{equation}
where $\sum_{ab}g_{Rab}[Re
P_R^{ab}(m_R^2)-P_R^{ab}(m^2)]=\Pi_R(m^2)=\Pi_{RR}(m^2)$ takes
into account the finite width corrections of the resonance which
are the one loop contribution to the self-energy of the $R$
resonance from the two-particle intermediate  $ab$ states. In the
$q^2\bar q^2$ model of the scalar particle and in the model of the
$K\bar K$ molecule the $f_0$ and $a_0$ mesons are strongly coupled
with the $K\bar K$ channel under threshold of which they are. The
ordinary resonance expression of the propagator, in view of it, is
changed considerably and the account of $\sum_{ab}g_{Rab}[Re
P_R^{ab} (m_R^2)-P_R^{ab}(m^2)]$  corrections is necessary.

For the pseudoscalar $ab$ mesons and $m_a\geq m_b,\ s>m_+^2$ one
has  \cite{charge,achasov-84,achasov-95}:
\begin{eqnarray}
\label{polarisator}
&&P^{ab}_R(m^2)=\frac{g_{Rab}}{16\pi}\left[\frac{m_+m_-}{\pi
m^2}\ln \frac{m_b}{m_a}+\right.\nonumber\\
&&\left.+\rho_{ab}\left(i+\frac{1}{\pi}\ln\frac{\sqrt{m^2-m_-^2}-
\sqrt{m^2-m_+^2}}{\sqrt{m^2-m_-^2}+\sqrt{m^2-m_+^2}}\right)\right]
\end{eqnarray}
For $m_-<m<m_+$
\begin{eqnarray}
&&P^{ab}_{R}(m^2)=\frac{g_{Rab}}{16\pi}\left[\frac{m_+m_-}{\pi
m^2}\ln \frac{m_b}{m_a}-|\rho_{ab}(m)|+\right.\nonumber\\
&&\left.+\frac{2}{\pi}|\rho_{ab}(m)
|\arctan\frac{\sqrt{m_+^2-m^2}}{\sqrt{m^2-m_-^2}}\right].
\end{eqnarray}
For $m<m_-$
\begin{eqnarray}
&&P^{ab}_{R}(m^2)=\frac{g_{Rab}}{16\pi}\left[\frac{m_+m_-}{\pi
m^2}\ln \frac{m_b}{m_a}-\right.\nonumber\\
&&\left.-\frac{1}{\pi}\rho_{ab}(m)\ln\frac{\sqrt{m_+^2-m^2}-
\sqrt{m_-^2-m^2}}{\sqrt{m_+^2-m^2}+\sqrt{m_-^2-m^2}}\right].
\end{eqnarray}
and
\begin{equation}
\label{rho-ab}
\rho_{ab}(m)=\sqrt{(1-\frac{m_+^2}{m^2})(1-\frac{m_-^2}{m^2})}\qquad
m_{\pm}=m_a\pm m_b
\end{equation}
The constants  $g_{Rab}$ are related to the width
\begin{equation}
\Gamma(R\to ab,m)=\frac{g_{Rab}^2}{16\pi m}\rho_{ab}(m).
\label{f0pipi}
\end{equation}

 The function $g(m^2)$ has  the
following expression \cite{achasov-89}.
 For $m<2m_{K^+}$

\begin{eqnarray}
&&g(m^2)=\frac{e}{2(2\pi)^2}g_{\phi K^+K^-}\Biggl\{
1+\frac{1-\rho^2(m^2)}{\rho(m^2_{\phi})^2-\rho(m^2)^2}\times\nonumber\\
&&\Biggl[2|\rho(m^2)|\arctan\frac{1}{|\rho(m^2)|}
-\rho(m^2_{\phi})\lambda(m^2_{\phi})+i\pi\rho(m^2_{\phi})-\nonumber\\
&&-(1-\rho^2(m^2_{\phi}))\Biggl(\frac{1}{4}(\pi+
i\lambda(m^2_{\phi}))^2- \nonumber\\
&&-\Biggl(\arctan\frac{1}{|\rho(m^2)|}\Biggr)^2
\Biggr)\Biggr]\Biggr\},
\end{eqnarray}
where
\begin{equation}
\rho(m^2)=\sqrt{1-\frac{4m_{K^+}^2}{m^2}}\qquad
\lambda(m^2)=\ln\frac{1+\rho(m^2)}{1-\rho(m^2)}.
\end{equation}

 For $m>2m_{K^+}$
\begin{eqnarray}
&&g(m^2)=\frac{e}{2(2\pi)^2}g_{\phi K^+K^-}\Biggl\{
1+\frac{1-\rho^2(m^2)}{\rho(m^2_{\phi})^2-\rho(m^2)^2}\times\nonumber\\
&&\times\Biggl[\rho(m^2)(\lambda(m^2)-i\pi)-
\rho(m^2_{\phi})(\lambda(m^2_{\phi})-i\pi)-\nonumber\\
&&\frac{1}{4}(1-\rho^2(m^2_{\phi}))
\Biggl((\pi+i\lambda(m^2_{\phi}))^2-
(\pi+i\lambda(m^2))^2\Biggr)\Biggr]\Biggr\}.
\end{eqnarray}
We use the constant $g_{\phi K^+K^-}=4.68$ related to the width by
following way:
\begin{equation}
\Gamma(\phi\to K^+K^-)=\frac{1}{3}\frac{g_{\phi
K^+K^-}^2}{16\pi}m_{\phi} \rho(m_{\phi}^2)^3
\end{equation}

\section{Results.}

\begin{figure}
 \centerline{\epsfxsize=12cm \epsfysize=8cm
\epsfbox{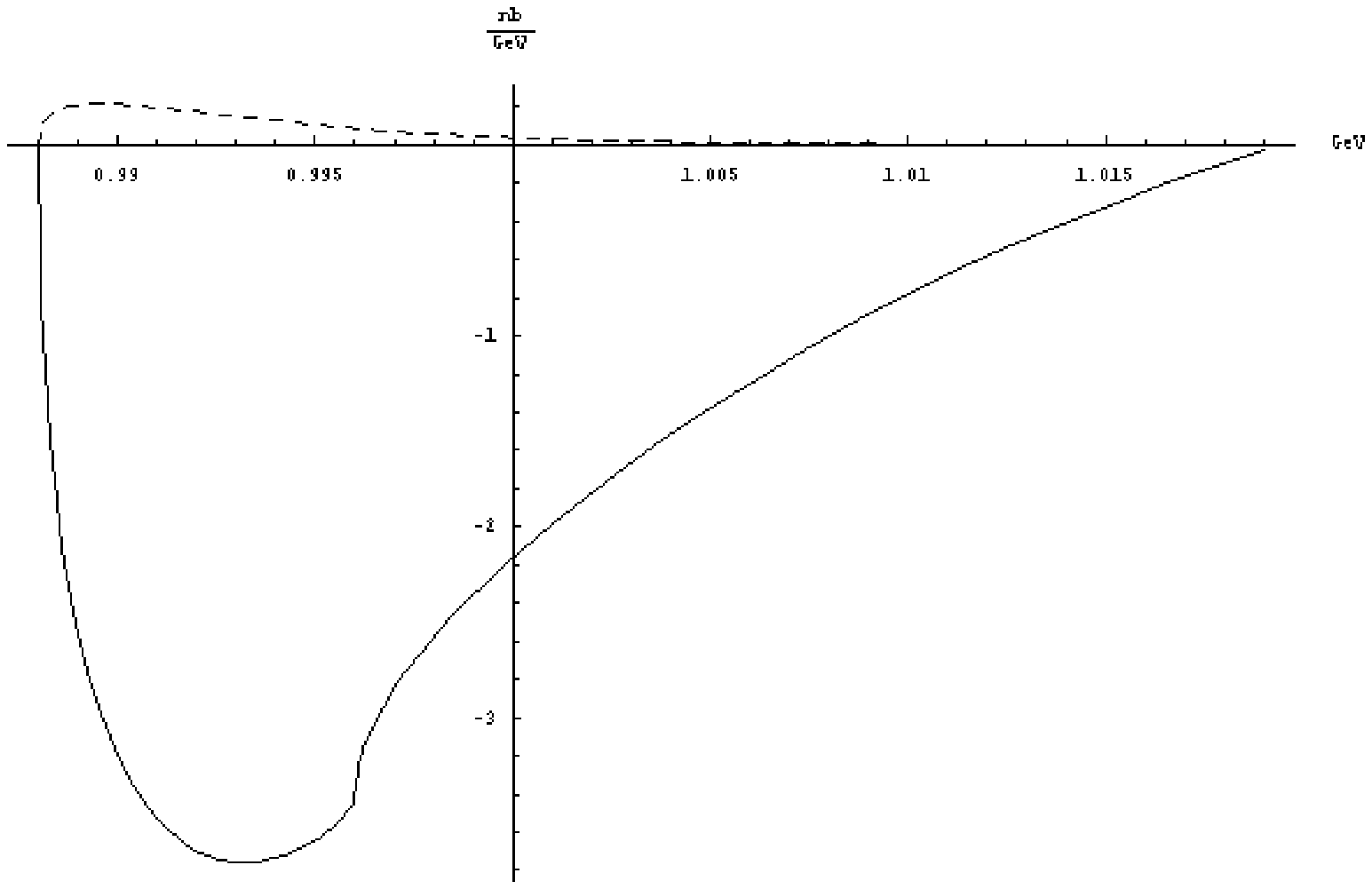}}
 \caption{ The solid line is the $e^+e^-\to\gamma K^+K^-$
 differential cross section $d\sigma_{\phi}/dm+d\sigma_{int}/dm$,
 see text. The dashed line is the differential cross section of the signal $d\sigma_{\phi}/dm$,
 see Eq.(\ref{ee-signal}).  The angle cuts are a=0.7, b=0.955. }
 \label{spectr1}
\end{figure}

\begin{figure}
 \centerline{\epsfxsize=12cm \epsfysize=8cm
\epsfbox{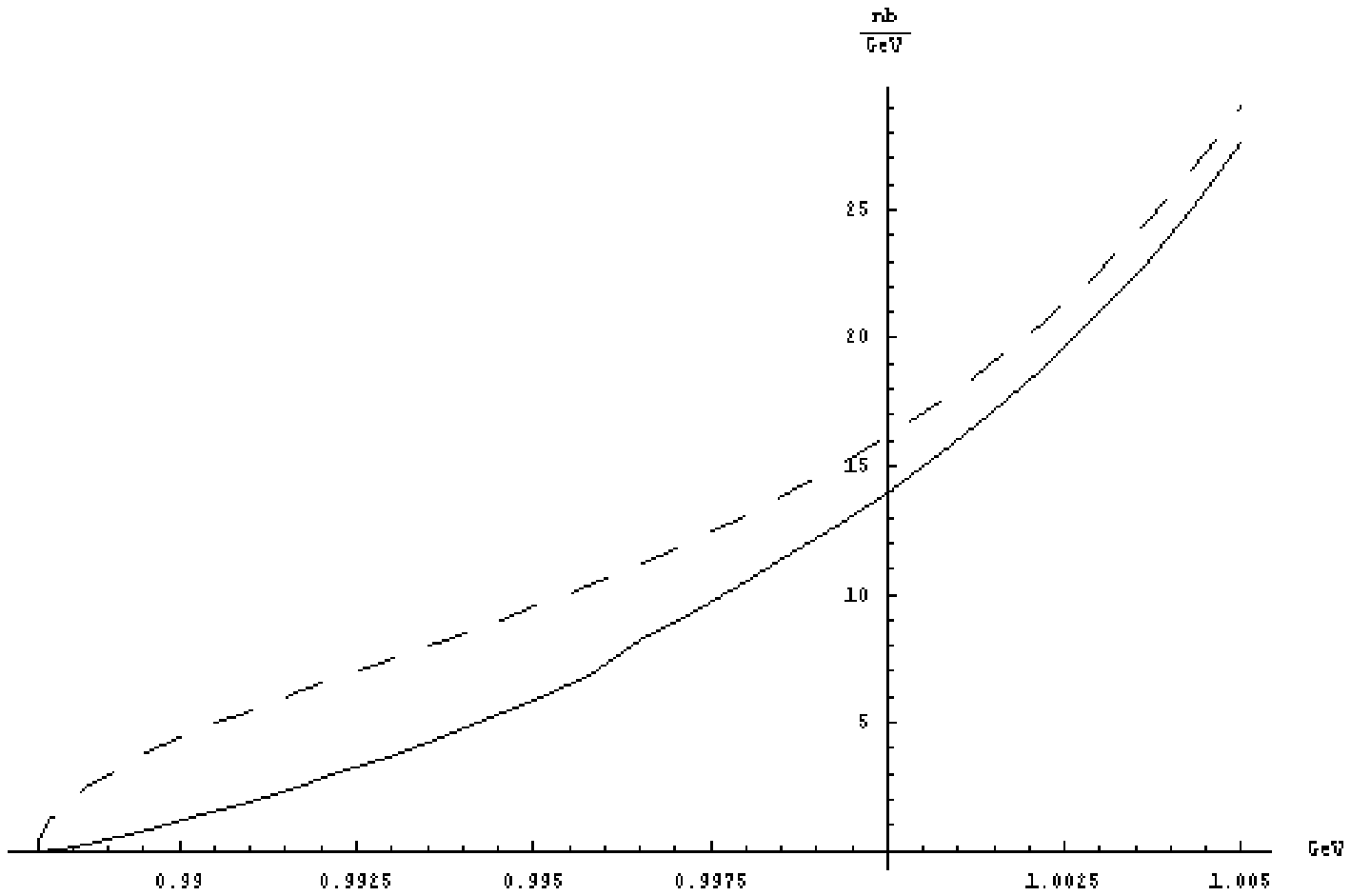}}
 \caption{ The solid line is the $e^+e^-\to\gamma K^+K^-$ differential cross section
 $d\sigma_{\phi}/dm+d\sigma_{int}/dm+d\sigma_{f}/dm+d\sigma_{i}/dm$, see text.
   The dashed line is the differential cross section of the
   background $d\sigma_{f}/dm+d\sigma_{i}/dm$, see text.
 The angle cuts are a=0.7, b=0.955. }
 \label{spectr11}
\end{figure}

To demonstrate we present the spectra for the $e^+e^-\to\gamma
K^+K^-$ reaction $d\sigma_{\phi}/dm+d\sigma_{int}/dm$ and
$d\sigma_{\phi}/dm+d\sigma_{int}/dm+d\sigma_{f}/dm+d\sigma_{i}/dm$
on Fig. \ref{spectr1} and Fig. \ref{spectr11}, using the constants
obtained from our fitting of the SND data \cite{achasov-01}:

\begin{eqnarray}
 &&g_{f_0K^+K^-}=4.021\pm0.011\  \mbox{GeV},\ \
g_{f_0\pi^0\pi^0}=1.494\pm0.021\ \mbox{GeV},\ \
m_{f_0}=0.996\pm0.0013\ \mbox{GeV}, \nonumber
\\ && g_{\sigma K^+K^-}=0, \ \ g_{\sigma\pi^0\pi^0}=2.58\pm0.02\ \mbox{GeV}, \ \
m_{\sigma}=1.505\pm0.012\ \mbox{GeV}, \nonumber
\\ &&C=0.622\pm0.04\  \mbox{GeV}^2,\ \
g_{f_0K^+K^-}^2/4\pi=1.29\pm0.017\  \mbox{GeV}^2.
\end{eqnarray}
and
\begin{eqnarray}
&&m_{a_0}=985.51\pm0.8\ \ \mbox{MeV} \nonumber \\
&&g_{a_0K^+K^-}=2.747\pm0.428\ \ \mbox{GeV};\
\frac{g_{a_0K^+K^-}^2}{4\pi}=0.6\pm0.015\ \ \mbox{GeV}^2
\end{eqnarray}

The branching ratio of the final state radiation decay
$\phi\to\gamma K^+K^-$ is $BR(\phi\to\gamma
K^+K^-,b=0.955,2m_K<m<1.01\ GeV)=9.41\cdot10^{-5}$. For the pure
signal we have $BR(\phi\to\gamma(a_0+f_0)\gamma
K^+K^-,b=0.955,2m_K<m<1.01\ GeV)=2.15\cdot10^{-6}$ and for the
signal plus interference term we have
$BR(\phi\to\gamma(a_0+f_0)\gamma K^+K^-,b=0.955,2m_K<m<1.01\
GeV)=-1.54\cdot10^{-5}$. For comparison, the branching ratio for
the $\phi\to\gamma(a_0+f_0)\to\gamma K^0\bar K^0$, for this set of
parameters is equal  $BR(\phi\to\gamma(a_0+f_0)\to\gamma K^0\bar
K^0,b=1.0,2m_{K_0}<m<m_{\phi}\ GeV)=4.36\cdot10^{-8}$.

\begin{figure}
 \centerline{\epsfxsize=12cm \epsfysize=8cm
\epsfbox{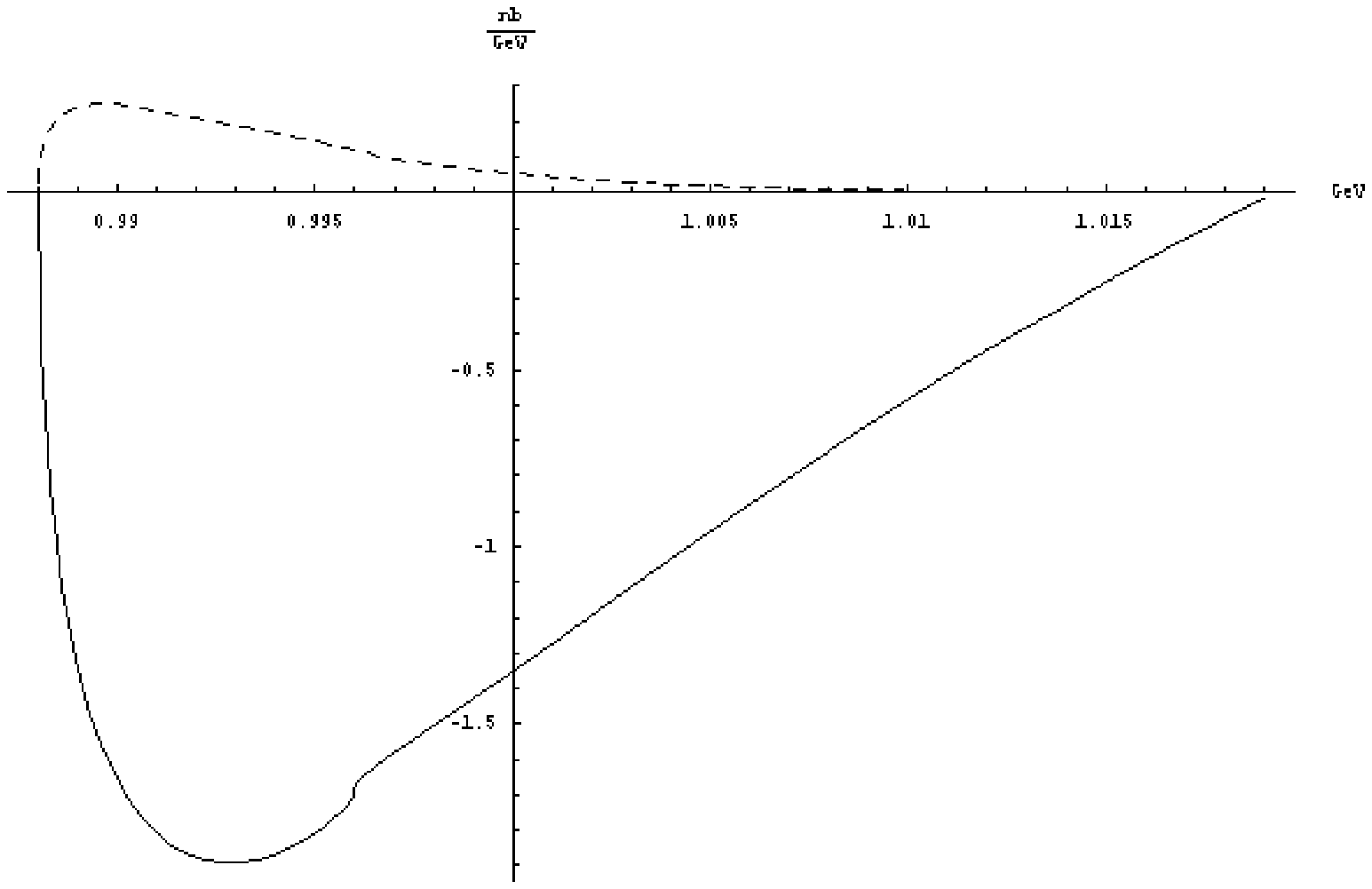}}
 \caption{  The solid line is the $e^+e^-\to\gamma K^+K^-$ differential
  cross section $d\sigma_{\phi}/dm+d\sigma_{int}/dm$,
 see text. The dashed line is the $e^+e^-\to\gamma K^+K^-$
 differential cross section of the signal $d\sigma_{\phi}/dm$,
 see Eq.(\ref{ee-signal}).  The angle cuts are a=0.7, b=0.955.}
 \label{spectr2}
\end{figure}

\begin{figure}
 \centerline{\epsfxsize=12cm \epsfysize=8cm
\epsfbox{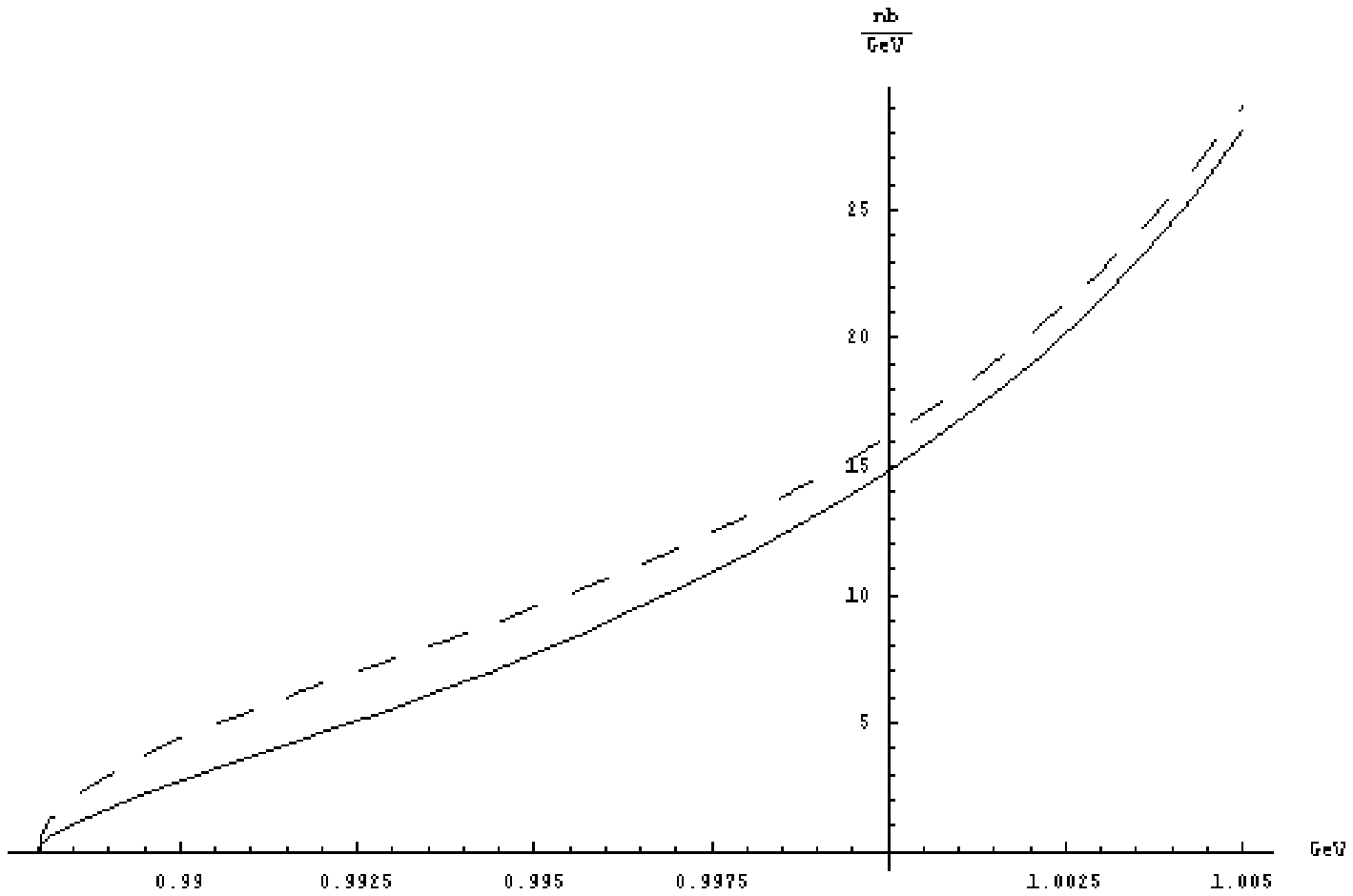}}
 \caption{ The solid line is the $e^+e^-\to\gamma K^+K^-$ differential cross section
 $d\sigma_{\phi}/dm+d\sigma_{int}/dm+d\sigma_{f}/dm+d\sigma_{i}/dm$, see text.
   The dashed line is the $e^+e^-\to\gamma K^+K^-$ differential cross section of the
   background $d\sigma_{f}/dm+d\sigma_{i}/dm$, see text.
 The angle cuts are a=0.7, b=0.955.}
 \label{spectr22}
\end{figure}

On  Fig. \ref{spectr2} and Fig. \ref{spectr22} we present the
variant of the SND data fitting \cite{snd-fit,snd-ivan} without
$f_0$ and $\sigma$ mixing

\begin{eqnarray}
 && m_{f_0}=0.9698\pm0.0045,\ \
 g_{f_0K^+K^-}^2/4\pi=2.47\pm^{0.73}_{0.51} \mbox{GeV}^2 \\ \nonumber
&&g_{f_0\pi^+\pi^-} ^2/4\pi=0.54\pm^{0.09}_{0.08} \mbox{GeV}^2,
\end{eqnarray}
and
\begin{eqnarray}
&&m_{a_0}=994\pm^{33}_{8}\ \ \mbox{MeV} \nonumber \\
&&\frac{g_{a_0K^+K^-}^2}{4\pi}=1.05\pm^{0.36}_{0.25}\ \
\mbox{GeV}^2
\end{eqnarray}

Analogously, for the pure signal we have
$BR(\phi\to\gamma(a_0+f_0)\to\gamma K^+K^-,b=0.955,2m_K<m<1.01\
GeV)=8.12\cdot10^{-7}$ and for the signal plus interference term
we have $BR(\phi\to\gamma(a_0+f_0)\to\gamma
K^+K^-,b=0.955,2m_K<m<1.01\ GeV)=-9.58\cdot10^{-6}$. The branching
ratio for the $\phi\to\gamma(a_0+f_0)\to\gamma K^0\bar K^0$, for
this set of parameters is equal
$BR(\phi\to\gamma(a_0+f_0)\to\gamma K^0\bar
K^0,b=1.0,2m_{K_0}<m<m_{\phi}\ GeV)=1.29\cdot10^{-8}$.

It is seen from our analysis  that the pictures are the same
qualitatively  for the both sets of parameters but the branching
ratio for the signal plus interference is predicted to within 1.5
and the branching ratio for the pure signal is predicted to within
2.5.

\section{Conclusion}
So, the analysis tells us that the research of the decays
$\phi\to\gamma K^+K^-$ in the processes $e^+e^-\to\phi\to\gamma
K^+K^-$ and   $\gamma N\to\phi N(\Delta)\to \gamma
K^+K^-N(\Delta)$ is the real problem that could be investigated in
Novosibirsk at VEPP-2000, in Frascati at DAFNE, and in Newport
News at Jefferson Laboratory. This research could give much
information about the inner structure of the $a_0$ and $f_0$
mesons and allow to check the hypothesis   of considerable $s\bar
s$ part in the $a_0$ and $f_0$ mesons.

\section{Acknowledgement}

We are thankful to V.V. Kulikov for a stimulating discussion. This
work was supported in part by INTAS-RFBR, grant IR-97-232.

\vspace{2cm}

\vspace{2cm}

\vspace{2cm}

\vspace{2cm}

\vspace{5cm}

\vspace{5cm}

\end{document}